\documentclass[aps,prl,showpacs,superscriptaddress,twocolumn]{revtex4}

\usepackage{mathrsfs}
\usepackage{amsfonts}
\usepackage{amssymb}
\usepackage{amsmath}
\usepackage{graphicx}
\usepackage{color}

\newcommand{\ket}[1]{\vert #1 \rangle}

\newcommand{\mean}[1]{\langle #1 \rangle}

\newcommand{\norm}[1]{\| #1 \|}
\newcommand{\abs}[1]{| #1 |}

\newcommand{\tr}{\mathrm{Tr}}

\newcommand{\Beq}{\begin{equation}}
\newcommand{\Eeq}{\end{equation}}
\newcommand{\BeqA}{\begin{eqnarray}}
\newcommand{\EeqA}{\end{eqnarray}}

\begin{document}

\title{Operator Quantum Zeno Effect: Protecting Quantum Information with Noisy Two-Qubit Interactions}

\author{Shu-Chao Wang}
\affiliation{Center for Quantum Technologies, National University of Singapore, 3 Science Drive 2, Singapore 117543, Singapore}
\affiliation{State Key Laboratory of Low Dimensional Quantum Physics, Department of Physics, Tsinghua University, Beijing 100084, People's Republic of China}

\author{Ying Li}
\email{ying.li.phys@gmail.com}
\affiliation{Center for Quantum Technologies, National University of Singapore, 3 Science Drive 2, Singapore 117543, Singapore}

\author{Xiang-Bin Wang}
\affiliation{State Key Laboratory of Low Dimensional Quantum Physics, Department of Physics, Tsinghua University, Beijing 100084, People's Republic of China}
\affiliation{Jinan Institute of Quantum Technology, Shandong Academy of Information and Communication Technology, Jinan 250101, People’s Republic of China}

\author{Leong Chuan Kwek}
\affiliation{Center for Quantum Technologies, National University of Singapore, 3 Science Drive 2, Singapore 117543, Singapore}
\affiliation{Institute of Advanced Studies, Nanyang Technological University,  60 Nanyang View Singapore 639673, Singapore}
\affiliation{National Institute of Education, 1 Nanyang Walk Singapore 637616, Singapore}

\date{\today}

\begin{abstract}
The time evolution of some quantum states can be slowed down or even stopped under frequent measurements.
This is the usual quantum Zeno effect.
Here, we report an operator quantum Zeno effect, in which the evolution of some physical observables is slowed down through measurements even though the quantum state changes randomly with time.
Based on the operator quantum Zeno effect, we show how we can protect quantum information from decoherence with two-qubit measurements, realizable with noisy two-qubit interactions.
\end{abstract}

\maketitle

\textit{Introduction}.---The quantum Zeno effect (QZE) predicts that frequent measurements can freeze the time evolution of a quantum state \cite{Beskow1967,Khalfin1968,Misra1977,Itano1990}.
Strictly speaking, the state need not be frozen in a single state, it could just be frozen within a multidimensional subspace, the Zeno subspace \cite{Facchi2002,Facchi2008}.
In a typical QZE, a set of observables commuting with each other are measured, and the state could evolve within the Zeno subspace in the presence of a Hamiltonian.
In this paper, we consider a different scenario for a quantum Zeno-like effect, where the observables need not be commutative.
In this case, the state may change with time as a result of the measurements.
In contrast to the usual QZE, we find that even though the state is not frozen in time, certain physical quantities can be frozen through frequent measurements.
We coin this effect as \textit{operator quantum Zeno effect} (OQZE).

Protecting quantum states from decoherence is crucial for practical quantum information processing.
A number of methods have been proposed for decoherence protection, including passive methods, e.g., decoherence-free subspace \cite{Palma1996,Zanardi1997,Lidar1998}, and active methods, e.g., quantum error correction code (QECC) \cite{Calderbank1996,Knill1997,Steane1996} and dynamical decoupling \cite{Viola1998,Zanardi1999,Viola1999}.
In the case of the two active methods, accurate quantum operations are required.
The QZE is also proposed for dealing with decoherence by frequently measuring stabilizers of QECCs \cite{Vaidman1996,Sarovar2005}.
Recently, it has been shown that QZE-based schemes can help suppress decoherence while allowing for full quantum control \cite{Paz-Silva2012}.
Compared with QECC and dynamical decoupling, QZE-based schemes tolerate erroneous measurements, because measurement outcomes are not read.
However, in previous QZE-based schemes, multiqubit measurements are required due to the nonlocality of stabilizers, which means one either needs multiqubit interactions or simulating multiqubit measurements with quantum circuits composed of single- and two-qubit quantum gates.

In this paper, we show a new protocol of suppressing decoherence based on the OQZE.
In our protocol, measurements are noncommutative providing protection with only single- and two-qubit measurements.
The measurements cause the state to evolve randomly even though the encoded logical states do not, provided the frequency of measurements is sufficiently high.
Since the measurement outcomes are not read, the operations of the measurements can be realized with noisy two-qubit interactions.

\textit{Operator quantum Zeno effect}.---Consider a set of measurements $\{ \mathcal{P}^{(k)} \}$ containing $K$ independent measurements, where $k=1,2,\ldots,K$.
Here, a measurement superoperator $\mathcal{P}^{(k)} \bullet = \sum _q M^{(k)}_q \bullet M^{(k)\dagger}_q$, and $\{ M^{(k)}_q \}$ satisfies the sum rule $\sum _q M^{(k)\dagger}_q M^{(k)}_q = \openone$.
Each measurement is performed instantly, and measurements are always sequentially done and denoted as $\mathcal{P} = \mathcal{P}^{(K)} \cdots \mathcal{P}^{(2)} \mathcal{P}^{(1)}$.
Suppose these measurements are performed $N$ times during the entire time of evolution $\tau$ at equal intervals.
For a system whose free evolution is governed by the Hamiltonian $H$, the superoperator describing the time evolution is $\mathcal{U}(t) = e^{\mathcal{L}t} $, where the generator $\mathcal{L} \bullet = -i[H , \bullet]$.
The state of the system at time $\tau$ is then given by
\begin{equation}
\rho(\tau) = [ \mathcal{U}(\tau/N) \mathcal{P} ]^N \rho(0),
\end{equation}
where $\rho(0)$ is the initial state.
On the other hand, the time evolution of an operator $A$ acting on the system is given by
\begin{equation}
A(\tau) = [ \mathcal{P}^{\dagger} \mathcal{U}(-\tau/N) ]^N A,
\end{equation}
so that $\tr[A(\tau) \rho(0)] = \tr[A \rho(\tau)]$ due to the cyclic property of the trace.
Here, $\mathcal{P}^{\dagger} = \mathcal{P}^{(1)\dagger} \mathcal{P}^{(2)\dagger} \cdots \mathcal{P}^{(K)\dagger}$ and $\mathcal{P}^{(k)\dagger} \bullet = \sum _j M^{(k)\dagger}_j \bullet M^{(k)}_j$.
Note that while $\mathcal{P}$ is a POVM, $\mathcal{P}^{\dagger}$ may not be one.

We now consider the case in which $A$ commutes with all measurements, i.e., $[A , M^{(k)}_q] =0$.
Expanding time evolution operators, $A(\tau) = \mathcal{V}A + O(1/N)$, where
\begin{equation}
\mathcal{V} = \{ \mathcal{P}^{\dagger} [ 1 - (\tau/N)\mathcal{L} ] \}^N,
\end{equation}
and $\tr[O(1/N) \rho]$ vanishes in the limit $N\rightarrow \infty$ for any state $\rho$ \cite{SupMat}, i.e., $\lim_{N\rightarrow \infty}\norm{O(1/N)}=0$, if $\norm{A}$ and $\norm{H}$ are both finite, where $\norm{\bullet}$ denotes the trace norm of a matrix.
A sufficient condition of the operator Zeno effect is that
\begin{equation}
\mathcal{P}^{\dagger} \mathcal{L} A = -i[\mathcal{P}^{\dagger}H , A] = 0.
\label{condition}
\end{equation}
Under this condition, the expansion of $\mathcal{V}A$ shows that $\mathcal{V}A = A$ \cite{SupMat}.
Therefore, $A(\tau)=A$ in the limit $N\rightarrow \infty$.

We would like to remark that the OQZE is different from a Heisenberg-picture formulation of the QZE.
The Heisenberg picture and the Schr\"{o}dinger picture are different formulations of the same physical process.
Under the Schr\"{o}dinger picture, the quantum state in a QZE is frozen by frequent measurements.
However, in the same picture, the state changes randomly under the measurements in the OQZE.

\textit{Time evolution of states}.--- Consider a sequence in which $\{ \mathcal{P}^{(k)} \}$ are all projective measurements of observables $\{ \Lambda^{(k)} \}$ for each $k$, respectively.
Clearly, $\mathcal{P}^{(k)}$ projects any state to an eigenstate of $\Lambda^{(k)}$.
So in this sequence, $\mathcal{P}^{(k)}$ projects an eigenstate of $\Lambda^{(k-1)}$ to an eigenstate of $\Lambda^{(k)}$.
If $\{ \Lambda^{(k)} \}$ do not commute with each other, these observables do not have common eigenstates, implying that the state evolve under the measurements even if the Hamiltonian of the system is switched off, i.e., $H=0$.
We reiterate that if $\{ \Lambda^{(k)} \}$ commute with each other, the state can evolve in a Zeno subspace due to a nonzero $H$.
This effect is typically known as the quantum Zeno dynamics \cite{Facchi2008}.
However, it is different from our OQZE.

In an OQZE, although the states may change, one can still employ the effect to protect quantum information without any feedback.
This is because if the condition Eq. (\ref{condition}) is satisfied by a set of operators $\{ A \}$ which defines a tensor-product subsystem, the states of this tensor-product subsystem can be frozen due to the Zeno effect.

\textit{Zeno quantum memory}.---We encode $m$ logical qubits using $n$ physical qubits.
In our protocol, the encoding need not be a QECC.
For $n$ qubits, $\Sigma = \{ \openone, X, Y, Z \} ^{\otimes n}$ is a subset of the Pauli group.
Elements of $\Sigma$ are all Hermitian and unitary, and any two elements of $\Sigma$ either commute or anticommute.
Logical qubits are represented by logical operators $L = \{ \overline{Z}_1, \overline{Z}_2, \ldots , \overline{Z}_m, \overline{X}_1, \overline{X}_2, \ldots , \overline{X}_m \}$, which is a subset of $\Sigma$.
Here, $\{ \overline{Z}_i, \overline{X}_i \}$ are Pauli operators of the $i$th logical qubit.
Logical operators satisfy $[\overline{Z}_i, \overline{Z}_j] = [\overline{X}_i, \overline{X}_j] = 0$ for all $i$ and $j$, $[\overline{Z}_i, \overline{X}_j] = 0$ for all $i \neq j$, and $\{\overline{Z}_i, \overline{X}_i \} = 0$ for all $i$.
The group $G(L)$, generated by $L$ and overall factors $\{ \pm 1,\pm i \}$, is a Pauli group of $m$ qubits.

Decoherence is induced by the Hamiltonian $H = H_S \otimes \openone_B + \openone_S \otimes H_B + H_{SB}$, where $H_S$, $H_B$, and $H_{SB}$ are Hamiltonians of the system, the bath, and the interaction between the system and the bath, respectively.
The Hamiltonian can be decomposed as $H = \sum_l a_l e_l$, where $E = \{ e_l \}$ is a subset of $\Sigma$, $\{ a_l \}$ are real coefficients or Hermitian operators of the bath, and $a_0$ is the coefficient of $\openone^{\otimes n}=\openone_S$.
We assume that $\norm{a_l}$ are all finite and $E \cap G(L) = \{ \openone^{\otimes n} \}$, otherwise logical qubits cannot be protected by our protocol.
Here, the second condition is automatically satisfied if $H$ is a \textbf{k}-local Hamiltonian (only for the system), and the locality of every element of $G(L)$ is higher than \textbf{k}, except $\openone^{\otimes n}$.

To protect logical qubits, elements of a subset of $\Sigma$, $C=\{c_k\}$, are measured sequentially.
The measurement superoperator corresponding to the element $c_k$ is $\mathcal{P}^{(k)} \bullet = P^{(k)}_+ \bullet P^{(k)}_+ + P^{(k)}_- \bullet P^{(k)}_-$, where $P^{(k)}_{\eta} = (\openone^{\otimes n} + \eta c_k)/2$.
These measurements satisfy the following conditions:
(i) elements of $C$ all commute with elements of $L$;
(ii) $G(C) \cap G(L) = \{ \openone^{\otimes n} \}$, where $G(C)$ is the group generated by $C$ and overall factors $\{ \pm 1,\pm i \}$;
and (iii) every element of $E$, except $\openone^{\otimes n}$, anticommutes with at least one element of $C$.
Conditions (i) and (ii) ensure that the measurements do not read out or destroy any information in the logical qubits, and that all elements of $G(L)$ commute with all $P^{(k)}_{\pm}$.
The condition (iii) results in $\mathcal{P}^{\dagger}H = a_0 \openone^{\otimes n}$ \cite{SupMat}, i.e., the sufficient condition for the OQZE in Eq. (\ref{condition}) is satisfied for all logical operators.
As a result of the Zeno effect, the evolution of logical operators can be frozen by frequent measurements, i.e., the stored quantum information is protected from decoherence.

\textit{Two-qubit measurements}.---One-local noise occurs if qubits are affected by the bath via two-local interactions.
We show that, if $G(C)$ is an Abelian group, two-qubit measurements are not enough to suppress general one-local noise.
For general one-local noise, $E$ contains all one-local elements of the system, i.e., $H = H_1 + H_{\text{others}}$, where $H_1 = \sum_{i=1}^n (a_{i,X}X_i + a_{i,Y}Y_i + a_{i,Z}Z_i)$.
To suppress noise on the qubit $i$, elements of $C$ must involve at least two of $\{X_i,Y_i,Z_i\}$.
Now, we suppose that $c_1$ and $c_2$ are two elements that involve $X_i$ and $Y_i$, respectively.
If $c_1$ and $c_2$ are both two-local, we write $c_1 = X_i \sigma$ and $c_2 = Y_i s$.
Because $[c_1,c_2]=0$, we have $\{\sigma,s\}=0$, which means that $\sigma$ and $s$ are operators of the same qubit and measurements of $c_1$ and $c_2$ projects two qubits, the qubit $i$ and the qubit corresponding to $\sigma$ and $s$, into a maximally entangled state.
We see that, these commutative two-qubit measurements project qubits into irrelevant maximally entangled pairs, i.e., the encoding of quantum information is not allowed.
Hence, commutative-measurement based protocols, or stabilizer-measurement based protocols, are not consistent with two-qubit measurements.

In our protocol, because $G(C)$ can be non-Abelian, we show that general one-local noise can be suppressed with only single- and two-qubit measurements.

\begin{figure}[tbp]
\includegraphics[width=8.0 cm]{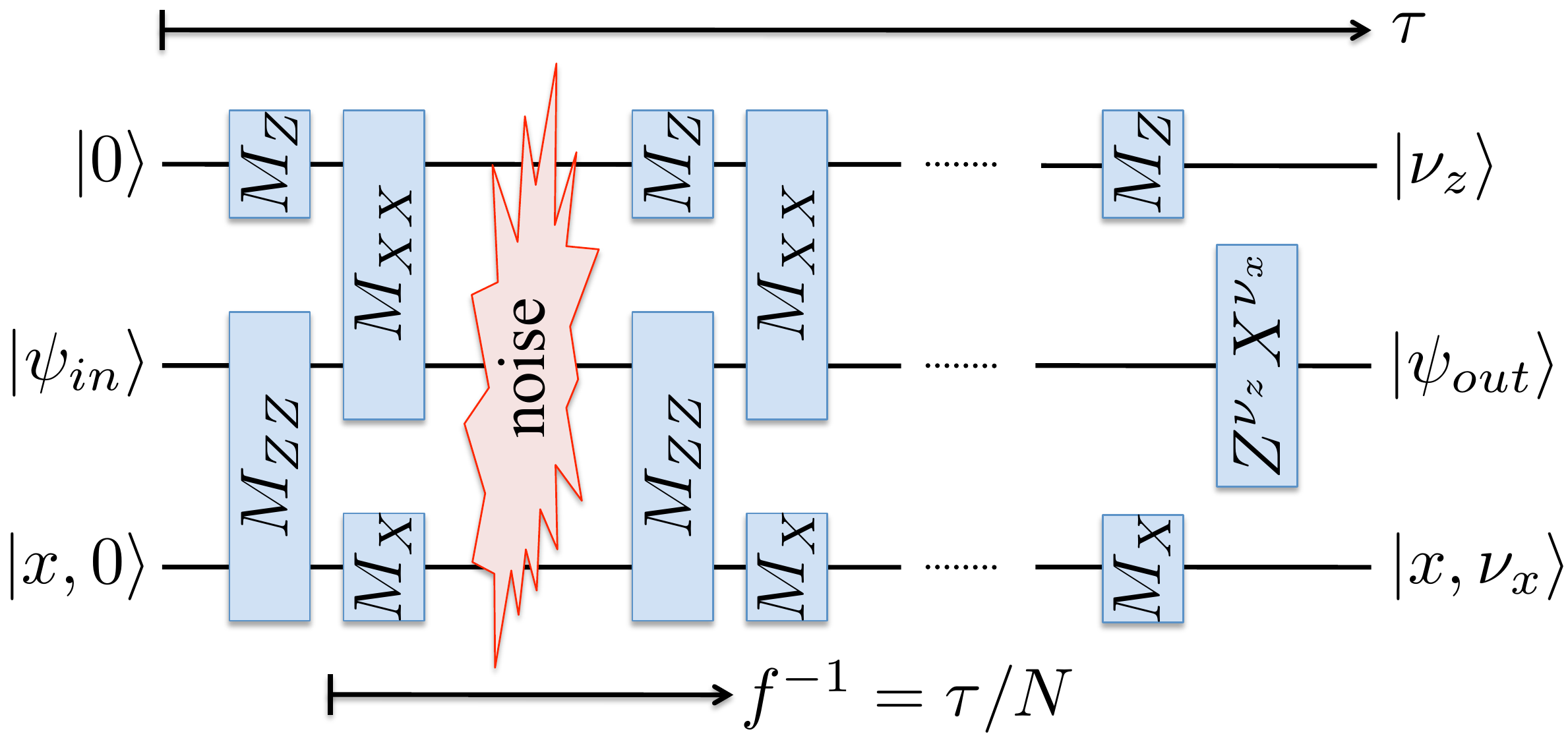}
\caption{
The scheme of protecting one logical qubit encoded in three qubits from noise.
The quantum state $\ket{\psi_{in}}$ is encoded into three qubits by initializing the other two qubits in states $\ket{0}$ and $\ket{x,0}$, respectively.
Here, $\ket{x,\nu}$ is an eigenstate of $X$ with eigenvalue $(-1)^{\nu}$.
To protect the logical state for time $\tau$, $N$ sets of measurements are performed with the frequency $f=N/\tau$.
Each set includes measurements of $Z$, $ZZ$, $X$, and $XX$ on corresponding qubits.
Here, we use $M_{\sigma}$ to denote the measurement of $\sigma$.
The logical qubit is decoded with $M_{Z}$ and $M_{X}$, whose outcomes are $\nu_Z$ and $\nu_X$, respectively.
Finally, a single-qubit operation is performed in order to correct the Pauli frame of the output state $\ket{\psi_{out}}$.
}
\label{circuit}
\end{figure}

\textit{Three-qubit encoding - an example}.---We consider three qubits as shown in Fig. \ref{circuit}.
Only one logical qubit is encoded as $\overline{Z} = Z_1Z_2$ and $\overline{X} = X_2X_3$.
The measurements include $C=\{ Z_1, Z_2Z_3, X_3, X_1X_2 \}$, in which $Z_1$ and $Z_2Z_3$ ($X_3$ and $X_1X_2$) are measured simultaneously.
$Z_2Z_3$ and $X_1X_2$ measurements are both two-qubit parity projections.
Initially, the logical quantum state is encoded in the logical computational basis $\ket{\overline{\mu}_{in}} = \ket{0}_1\ket{\mu}_2\ket{x,0}_3$, where $\mu=0,1$.
Here, $\ket{x,\mu}_3 = (1/\sqrt{2}) [\ket{0}_3+(-1)^{\mu}\ket{1}_3]$ are eigenstates of $X_3$.
During the process, the basis states used for encoding can change randomly, depending on the outcomes $\{\nu_z,\nu_{zz},\nu_x,\nu_{xx}\}$ of the measurements $C$ respectively.
Here, $\nu_k = 0,1$ are corresponding to eigenvalues $1,-1$, respectively.
After measuring $Z_1$ and $Z_2Z_3$, the basis states are
\begin{equation}
\ket{\overline{\mu}_z} = \ket{\nu_z}_1\ket{\mu \oplus \nu_z}_2
\ket{\mu \oplus \nu_z \oplus \nu_{zz}}_3.
\end{equation}
After measuring $X_3$ and $X_1X_2$, the basis states are
\begin{equation}
\ket{\overline{\mu}_x} = (-1)^{\mu (\nu_x+\nu_{xx})}
\ket{\phi_{\mu,\nu_{xx}}}_{1,2} \ket{x,\nu_x}_3,
\end{equation}
where the Bell states
\begin{equation}
\ket{\phi_{\mu,\nu_{xx}}}_{1,2} =
\frac{1}{\sqrt{2}} [\ket{\mu}_1\ket{0}_2+(-1)^{\nu_{xx}}\ket{1 \oplus \mu}_1\ket{1}_2].
\end{equation}
Even basis states change randomly, one does not have to record the measurement outcomes during the process.
To read out the logical qubit, $Z_1$ and $X_3$ are measured simultaneously, and only these two outcomes are recorded.
The logical operators are converted into single-qubit operators as $\overline{Z} = (-1)^{\nu_z} Z_2$ and $\overline{X} = (-1)^{\nu_x} X_2$, and basis states $\ket{\overline{\mu}_{out}} = (-1)^{\mu\nu_x}\ket{\nu_z}_1\ket{\mu \oplus \nu_z}_2\ket{x,\nu_x}_3$, which only depend on the last two measurement outcomes.

The time evolution of quantum logical operators is frozen by frequent measurements, i.e., their average values do not change.
Because quantum states of a qubit can always be described with the expression $\rho = \openone/2 + \mean{X}X + \mean{Y}Y + \mean{Z}Z$, where $\mean{\bullet}$ denotes the average value of $\bullet$ in the state $\rho$, we  conclude that the logical state has not evolved throughout the entire process.
It is also shown in Ref. \cite{SupMat} how the encoded quantum information is stabilized by the measurements.

\begin{figure}[tbp]
\includegraphics[width=8.0 cm]{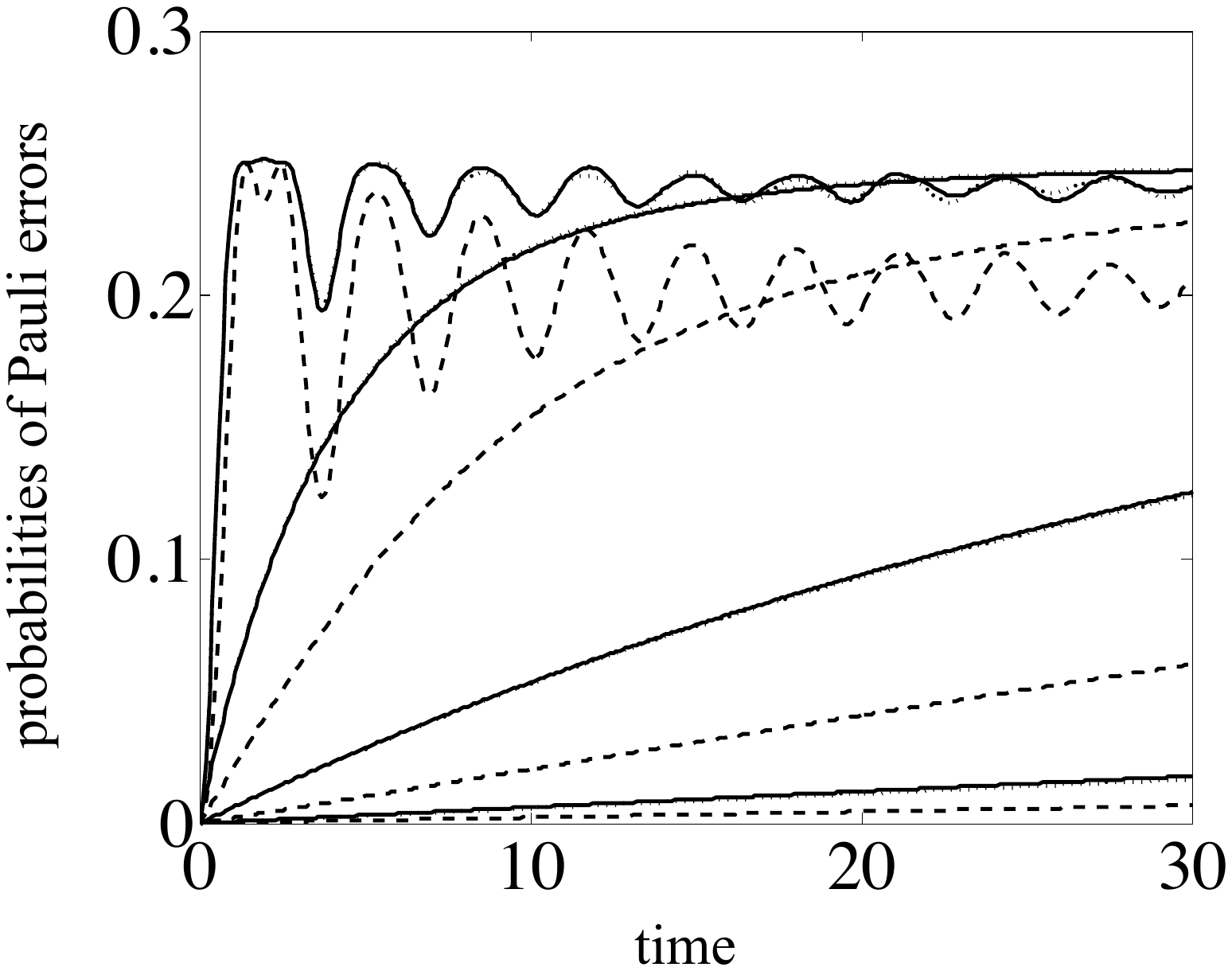}
\caption{
The probabilities of Pauli errors, $p_X$, $p_Y$, and $p_Z$ (solid line, dashed line, and dotted line, respectively).
Here, we consider one-local noise with parameters $\{\textbf{a}_i=(a_{i,X},a_{i,Y},a_{i,Z})\}$, which are uniformly distributed random vectors with $\norm{\textbf{a}_i}\leq a$.
The unit of time is $a^{-1}$.
One can find that $p_X$ and $p_Z$ are coincident.
By increasing the measurement frequency (from top to bottom: without measurements, $f=10$, $f=100$ and $f=1000$), one can reduce the probabilities of getting Pauli errors.
}
\label{error}
\end{figure}

\textit{Pauli errors}.---We quantitatively describe the performance of the Zeno quantum memory with error superoperators.
For any initial logical state $\rho_{in}$, the output logical state can always be written as $\rho_{out} = \mathcal{E}\rho_{in}$, where the error superoperator $\mathcal{E}$ is independent of the initial logical state.
In our three-qubit example, we find that $\mathcal{E}\bullet = F\bullet + p_X X\bullet X + p_Y Y\bullet Y + p_Z Z\bullet Z$ if noise is isotropic.
Here, $F=1-p_X-p_Y-p_Z$ is the fidelity of the quantum memory and $p_{\sigma}$ is the probability of the Pauli error $[\sigma]$, where $\sigma=X,Y,Z$.
In Fig. \ref{error}, we show error probabilities changing with the storage time for varying measurement frequencies.
One can find that error probabilities can be reduced by increasing measurement frequencies.

\begin{figure}[tbp]
\includegraphics[width=8.0 cm]{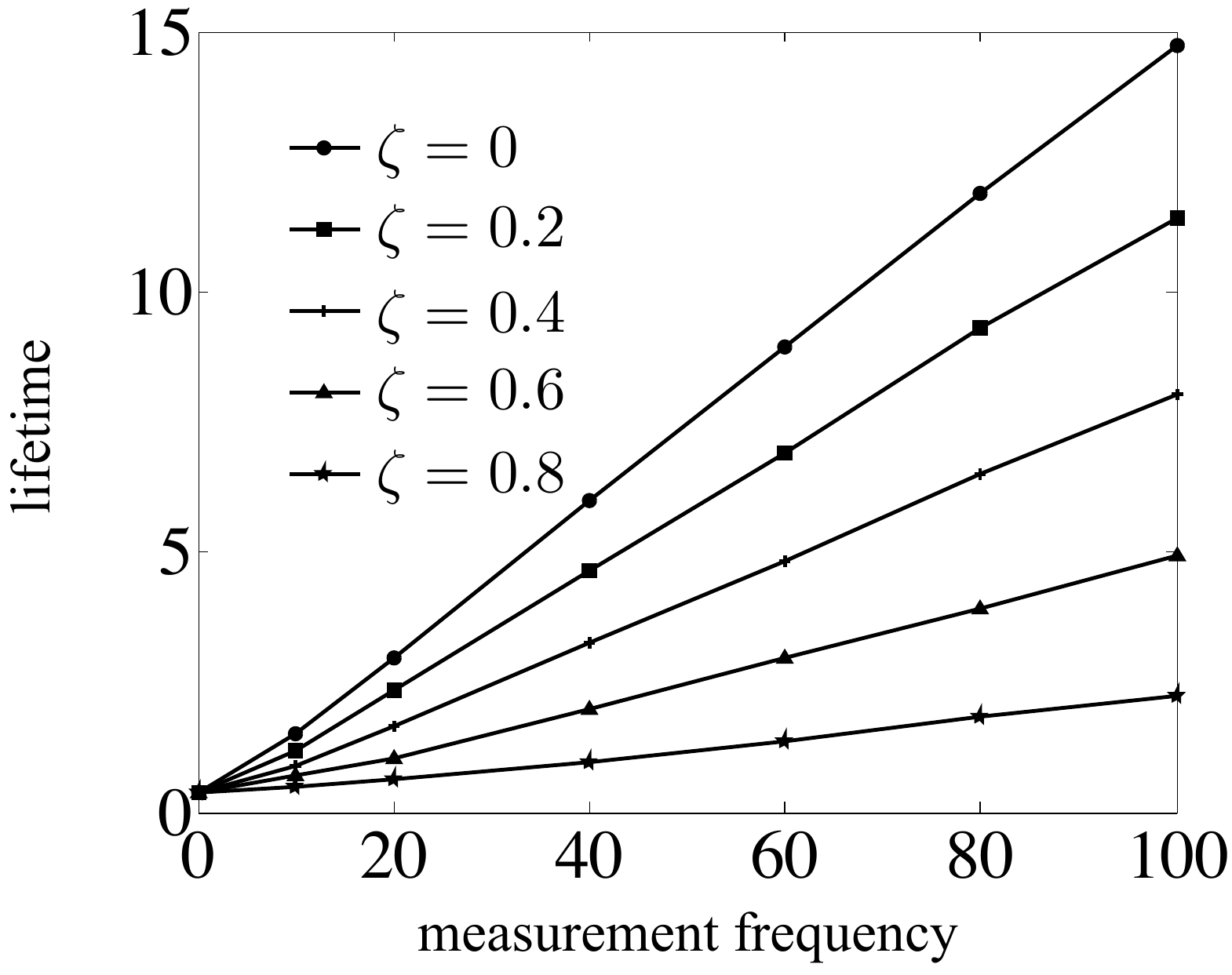}
\caption{
The lifetime of a logical qubit protected by both frequent measurements and the surface code.
Here, we utilize measurements described by superoperators $\mathcal{P}_{\sigma}(\zeta)\bullet =[(1+\zeta)/2]\bullet + [(1-\zeta)/2]\sigma \bullet \sigma$, where $\sigma$ is the measured operator.
When $\zeta =0$ the superoperator is a projective measurement investigated in the text.
When $\zeta >0$, the superoperator corresponds to a weak measurement.
By increasing the measurement frequency, one can extend the lifetime of the logical qubit.
Numerical results show that our scheme also works with weak measurements.
}
\label{lifetime}
\end{figure}

\textit{Decoherence time}.---The Zeno quantum memory can be combined with QECCs \cite{Calderbank1996,Knill1997,Steane1996}.
We propose to encode a high-level logical qubit in many low-level logical qubits stored in Zeno memories.
Here, we take Kitaev's surface code as an example \cite{Dennis2002}.
A surface code quantum memory is robust against errors with a threshold of $\max \{p_X+p_Y,p_Z+p_Y\}<0.104$, if quantum gates are perfect.
The threshold of fault-tolerant quantum computing (FTQC) based on the surface code is $\sim 1\%$ \cite{Fowler2009,Wang2011}.
A gate error rate $1$ or $2$ orders of magnitude below the FTQC threshold will not significantly change the threshold of the quantum memory.
With error probabilities lower than this threshold, the fidelity of the surface-code logical qubit is always higher by encoding with more low-level qubits.
Thus, we can define a lifetime of Zeno quantum memories.
If the storing time is shorter than the lifetime, errors can be corrected by the surface code and the fidelity of the high-level logical qubit can be arbitrarily approaching unity with sufficient low-level logical qubits.
The lifetime is not the maximum time of storing high-level logical qubit but the time before active error corrections start.
In Fig. \ref{lifetime}, we show the lifetime of the three-qubit Zeno quantum memory.

\textit{Noisy two-qubit interaction}.---In our three-qubit Zeno quantum memory protocol, because outcomes are not recorded, parity projections can be realized with noisy Ising interactions.
The parity projection superoperator can be rewritten as $\mathcal{P}_{\sigma\sigma}\bullet = (1/2)\bullet + (1/2)\sigma\sigma\bullet\sigma\sigma$, where $\sigma=X,Z$.
Therefore, the parity projection is equivalent to randomly performing the identity operation or the $\sigma\sigma$ operation with the same probability $1/2$.
We describe the time evolution driven by a noisy Ising interaction as $\mathcal{U}_{\sigma\sigma}\bullet = \int dJp(J)e^{-iJt\sigma\sigma}\bullet e^{iJt\sigma\sigma}$, where $p(J)$ is the probability density of the coupling constant $J$.
Here, we choose a time $t$ satisfying $\int dJp(J)\sin(Jt)\cos(Jt)=0$ and $p_{\sigma\sigma}=\int dJp(J)\sin^2(Jt)>1/2$.
Then one can realize the parity projection by randomly performing the noise evolution with the probability $1/(2p_{\sigma\sigma})$.

\textit{Conclusion}.---In this paper, we investigated the theory of the OQZE, in which states may evolve under frequent measurements even though the time evolution of certain operators can be frozen.
We find a sufficient condition for the OQZE, though we believe that this condition is not a necessary condition and a more general condition may exist.
The OQZE can be used to protect quantum information stored in logical qubits from decoherence.
By taking advantage of the OQZE, quantum information can be protected with two-qubit measurements.
We have only considered projective measurements analytically, but our numerical results show that our protocol also works with weak measurements \cite{Peres1990,Paz-Silva2012}, as shown in Fig. \ref{lifetime}.
Two-qubit measurements, and even many-qubit measurements can be implemented (even fault tolerantly) using local operations on the platform of quantum computers \cite{Paz-Silva2012}.
As shown in this paper, two-qubit measurements can also be simulated with noisy Ising interactions.
Though we only show that the three-qubit encoding can protect the quantum information from general one-qubit noise, we believe that multiqubit noise can be corrected by encoding each logical qubit into more physical qubits.
Finally, we note that we have not considered the feasibility of using quantum control during the protection of the logical qubits in our protocol, which deserves future investigation.

\begin{acknowledgments}
S.-C.W., Y.L. and L.C.K. acknowledge support from the National Research Foundation \& Ministry of Education, Singapore.
X.-B.W. and S.-C.W. acknowledge the support from the 10000-Plan of Shandong province, the National High-Tech Program of China Grants No. 2011AA010800 and No. 2011AA010803 and NSFC Grants No. 11174177 and No. 60725416.
\end{acknowledgments}

%%%%%%%%%%%%%%%%%%%%%%%%%%%%%
\newpage
\widetext

\begin{center}
{\huge Supplementary Material}
\end{center}

\section{The expansion of evolution superoperators}

The evolution of an operator is given by Eq. (2) in the main text.
An expansion of the evolution superoperator gives
\begin{equation}
A(\tau) = \{ \mathcal{P}^{\dagger} [ 1 - (\tau/N)\mathcal{L} + \sum_{l=2}^{\infty} \frac{(-\tau/N)^l}{l!}\mathcal{L}^l ] \}^N A.
\end{equation}
We rewrite $A(\tau)$ as two parts $A(\tau) = \mathcal{V}A + O(1/N)$, where
\begin{equation}
\mathcal{V}A = \{ \mathcal{P}^{\dagger} [ 1 - (\tau/N)\mathcal{L} ] \}^N A
\end{equation}
and
\begin{equation}
O(1/N) = A(\tau) - \mathcal{V}A = \sum_{m=0}^{N-1} ]
\{ \mathcal{P}^{\dagger} [ 1 - (\tau/N)\mathcal{L} ] \}^{N-1-m}
\frac{(-\tau/N)^2}{2!}\mathcal{L}^2
\{ \mathcal{P}^{\dagger} [ 1 - (\tau/N)\mathcal{L} ] \}^m
+\cdots .
\end{equation}

A further expansion of $\mathcal{V}A$ can be written as
\begin{equation}
\mathcal{V}A= \sum_{n=0}^N (-\tau/N)^n \sum_{\{m_0,m_1,\ldots ,m_n\}}
\mathcal{P}^{\dagger m_n} \mathcal{L} \cdots
\mathcal{P}^{\dagger m_2} \mathcal{L} \mathcal{P}^{\dagger m_1} \mathcal{L} \mathcal{P}^{\dagger m_0} A.
\end{equation}
Here, $\{m_0,m_1,\ldots ,m_n\}$ are all integers satisfying $\sum_{i=0}^n m_i = N$, in which $m_0$ is non-negative while others are all positive.
Because $[A , M^{(k)}_j] =0$, we have $\mathcal{P}^{\dagger m_0} A = A$ and
\begin{equation}
\mathcal{P}^{\dagger m_n} \mathcal{L} \cdots
\mathcal{P}^{\dagger m_2} \mathcal{L} \mathcal{P}^{\dagger m_1} \mathcal{L} \mathcal{P}^{\dagger m_0} A
= \mathcal{P}^{\dagger m_n} \mathcal{L} \cdots
\mathcal{P}^{\dagger m_2} \mathcal{L} \mathcal{P}^{\dagger m_1-1} \mathcal{P}^{\dagger} \mathcal{L} A.
\end{equation}
Therefore, if $\mathcal{P}^{\dagger} \mathcal{L} A = 0$, terms with a positive $n$ are all zero and $\mathcal{V}A = A$ as a result of the term with $n=0$.

Here, we prove that for any state $\rho$, $\lim_{N\rightarrow \infty} \tr[ O(1/N) \rho ] = 0$.
We begin with
\begin{equation}
\tr[ O(1/N) \rho ] =
\tr \left[ A \left( \{ [ 1 + (\tau/N)\mathcal{L} + \sum_{l=2}^{\infty} \frac{(\tau/N)^l}{l!}\mathcal{L}^l ] \mathcal{P} \}^N
- \{ [ 1 + (\tau/N)\mathcal{L} ] \mathcal{P} \}^N \right) \rho \right],
\label{traceO}
\end{equation}
which can be rewritten as
\begin{equation}
\tr[ O(1/N) \rho ] =
\sum_{\{n_i\}} \sum_{\{m_i\}} \alpha_{\{n_i\}\{m_i\}}
\tr( A \mathcal{P}^{m_N} \mathcal{L}^{n_N} \cdots
\mathcal{P}^{m_2} \mathcal{L}^{n_2} \mathcal{P}^{m_1} \mathcal{L}^{n_1} \mathcal{P}^{m_0} \rho ),
\end{equation}
where $\{n_i\}$ and $\{m_i\}$ are all non-negative integers and $\{ \alpha_{\{n_i\}\{m_i\}} \}$ are all non-negative real coefficients.
Because $\{\cdots \mathcal{P}^{m_2} \mathcal{L}^{n_2} \mathcal{P}^{m_1} \mathcal{L}^{n_1} \mathcal{P}^{m_0} \rho\}$ are all Hermitian operators, and a POVM does not increase the trace norm of a Hermitian operator (see the following two subsections for explanation), we have
\begin{equation}
\tr[ O(1/N) \rho ] \leq
\sum_{\{n_i\}} \sum_{\{m_i\}} \alpha_{\{n_i\}\{m_i\}}
(2\norm{H})^{\sum_i n_i} \norm{A},
\end{equation}
where right side can be obtained by replacing $\mathcal{P}$ with $1$, $A$ with $\norm{A}$, and $\mathcal{L}$ with $2\norm{H}$ in the right side of Eq. (\ref{traceO}), i.e.,
\begin{equation}
\{ e^{2\norm{H}\tau} - [ 1 + (\tau/N)(2\norm{H}) ]^N \} \norm{A}
= \sum_{\{n_i\}} \sum_{\{m_i\}} \alpha_{\{n_i\}\{m_i\}}
(2\norm{H})^{\sum_i n_i} \norm{A}.
\end{equation}
Because
\begin{equation}
\lim_{N\rightarrow \infty} [ 1 + (\tau/N)(2\norm{H}) ]^N = e^{2\norm{H}\tau},
\end{equation}
we finally obtain
\begin{equation}
\lim_{N\rightarrow \infty} \tr[ O(1/N) \rho ] = 0.
\end{equation}

\subsection*{Trace norm and Trace}

For two Hermitian operators $A$ and $B$, $\tr( AB ) = \tr( AB+BA )/2$, where $AB+BA$ is also a Hermitian operator.
For a Hermitian operator, the trace norm is the sum of the absolute values of eigenvalues.
Therefore, $\abs{ \tr( AB ) } = \abs{ \tr(AB+BA)/2 } \leq \norm{AB+BA}/2 \leq \norm{A}\norm{B}$.

\subsection*{Trace norm and POVM}

A Hermitian operator $A$, can be decomposed as $A = A_+ - A_-$, where $A_+$ and $A_-$ are two positive Hermitian operators corresponding to positive eigenvalues and negative eigenvalues of $A$, respectively.
Then, $\norm{A} = \tr (A_+ + A_-) $
For positive Hermitian operators, $\norm{A_\pm}=\tr A_\pm$.
Because $\mathcal{P}A_\pm$ are also positive Hermitian operators, $\norm{\mathcal{P}A} \leq \norm{\mathcal{P}A_+} + \norm{\mathcal{P}A_-} = \tr [\mathcal{P}(A_+ + A_-)]  = \norm{A}$.

\section{The condition (iii) of the Zeno quantum memory \& The condition of the operator Zeno effect}

Here we prove that the condition (iii) of the \textit{Zeno quantum memory} provides a sufficient condition of the operator Zeno effect, i.e., Eq. (4) in the main text.

As shown in the main text, the $k$th measurement superoperator is $\mathcal{P}^{(k)} \bullet = P^{(k)}_+ \bullet P^{(k)}_+ + P^{(k)}_- \bullet P^{(k)}_-$, where $P^{(k)}_{\eta} = (\openone^{\otimes n} + \eta c_k)/2$ and $\eta = \pm 1$.
We can rewrite the measurement superoperator as $\mathcal{P}^{(k)} \bullet = ( \bullet + c_k \bullet c_k )/2$.
Because $P^{(k)}_{\eta}$ are all Hermitian, $\mathcal{P}^{(k)\dagger} = \mathcal{P}^{(k)}$.

We assume that
\begin{equation}
\mathcal{P}^{(k)\dagger} \cdots \mathcal{P}^{(K)\dagger}H = \sum_l a_l^{(k)} e_l.
\end{equation}
Then,
\begin{equation}
\mathcal{P}^{(k+1)\dagger} \mathcal{P}^{(k)\dagger} \cdots \mathcal{P}^{(K)\dagger}H
= \sum_l a_l^{(k)} \mathcal{P}^{(k+1)\dagger} e_l,
\end{equation}
where
\begin{equation}
\mathcal{P}^{(k+1)\dagger} e_l = \frac{1}{2} ( e_l + c_{k+1}e_lc_{k+1} ).
\end{equation}
Here, $c_{k+1}$ either commutes or anticommutes with $e_l$, because they are both Hermitian elements of the Pauli group.
If $c_{k+1}$ commutes with $e_l$, $\mathcal{P}^{(k+1)\dagger} e_l = e_l$ and $a_l^{(k+1)} = a_l^{(k)}$ because of $c_{k+1}^2 = \openone^{\otimes n}$.
Similarly, if $c_{k+1}$ anticommutes with $e_l$, $\mathcal{P}^{(k+1)\dagger} e_l = 0$ and $a_l^{(k+1)} = 0$.

Therefore, if every element of $E=\{e_l\}$, except $\openone^{\otimes n}$, anticommutes with at least one element of $C=\{c_k\}$, $\mathcal{P}^{\dagger}H = H^{(K)} = a_0 \openone^{\otimes n}$, which results in $-i[\mathcal{P}^{\dagger}H , A] = 0$ for any operator $A$.

\section{Three-qubit encoding: the time evolution with noise}

In order to show how the encoded quantum information is stabilized by the measurements, we suppose that the state of the encoded logical qubit is the eigenstate of the logical operator $L= \alpha\overline{X} +\beta \overline{Y} +\gamma \overline{Z}$ with eigenvalue $+1$.
Here, $\overline{X} = X_2X_3$, $\overline{Y} = Z_1Y_2X_3$, $\overline{Z} = Z_1Z_2$ are logical Pauli operators, and $\alpha,\beta,\gamma$ are all positive real number satisfying $\alpha^2 + \beta^2 + \gamma^2 =1$.
We would like to remark that the average value of $(1+L)/2$, the projector corresponding to the stored quantum state, is the fidelity of the quantum memory.

After the first set of measurements, the state of three physical logical qubits is $\rho _0= (1+L)/8$ because four possible outcomes of $\{ \nu_{x},\nu_{xx} \}$ occur with the same probabilities.
Then, after the time evolution driven by the noise Hamiltonian and the second set of measurements, we get
\begin{eqnarray}
\rho _1 = \mathcal{P} e^{-iH\tau/N}\rho _0e^{iH\tau/N} = \mathcal{P} \rho _0 -i\frac{\tau}{N}\mathcal{P}[H,\rho _0]
- \frac{\tau^2}{2N^2}\mathcal{P}[H,[H,\rho _0]] + \cdots .
\end{eqnarray}
Because $Z_1$, $Z_2Z_3$, $X_1X_2$ and $X_3$ commute with $\overline{X}$, $\overline{Y}$ and $\overline{Z}$,
\begin{eqnarray}
\rho _1 = \rho _0 -i\frac{\tau}{N}[\mathcal{P}H,\rho _0]
- \frac{\tau^2}{2N^2}\mathcal{P}[H,[H,\rho _0]] + \cdots .
\end{eqnarray}
As a result of the condition (iii) of the \textit{Zeno quantum memory}, we have $\mathcal{P}H \propto \openone$ (the proof is similar to it of $\mathcal{P}^{\dagger}$).
Therefore, after the second set of measurements, the probability of errors $\sim 1/N^2$.
The effects of the subsequent evolutions under noise and measurements are similar.
And after $N$ sets of measurements, the final probability of errors $\sim 1/N$.
Therefore, by increasing $N$, one can reduce the probability of errors and stabilize the stored quantum information.


\begin{thebibliography}{9}

\bibitem{Beskow1967} A. Beskow and J. Nilsson, Ark. Fys. \textbf{34}, 561 (1967).

\bibitem{Khalfin1968} L. A. Khalfin, JETP Lett. \textbf{8}, 65 (1968).

\bibitem{Misra1977} B. Misra and E. C. G. Sudarshan, J. Math. Phys. \textbf{18}, 756 (1977).

\bibitem{Itano1990} W. M. Itano, D. J. Heinzen, J. J. Bollinger, and D. J. Wineland, Phys. Rev. A \textbf{41}, 2295 (1990).

\bibitem{Facchi2002} P. Facchi and S. Pascazio, Phys. Rev. Lett. \textbf{89}, 080401 (2002).

\bibitem{Facchi2008} P. Facchi and S. Pascazio, J. Phys. A \textbf{41}, 493001 (2008).

\bibitem{Palma1996} G. M. Palma, K.-A. Suominen, and A. K. Ekert, Proc. R. Soc. Lond. A \textbf{452}, 567 (1996).

\bibitem{Zanardi1997} P. Zanardi and M. Rasetti, Phys. Rev. Lett. \textbf{79}, 3306 (1997).

\bibitem{Lidar1998} D. A. Lidar, I. L. Chuang, and K. B. Whaley, Phys. Rev. Lett. \textbf{81}, 2594 (1998).

\bibitem{Calderbank1996} A. R. Calderbank and P. W. Shor, Phys. Rev. A \textbf{54}, 1098 (1996).

\bibitem{Knill1997} E. Knill and R. Laflamme, Phys. Rev. A \textbf{55}, 900 (1997).

\bibitem{Steane1996} A. M. Steane, Phys. Rev. Lett. \textbf{77}, 793 (1996).

\bibitem{Viola1998} L. Viola and S. Lloyd, Phys. Rev. A \textbf{58}, 2733 (1998).

\bibitem{Zanardi1999} P. Zanardi, Phys. Lett. A \textbf{258}, 77 (1999).

\bibitem{Viola1999} L. Viola, E. Knill, and S. Lloyd, Phys. Rev. Lett. \textbf{82}, 2417 (1999).

\bibitem{Vaidman1996} L. Vaidman, L. Goldenberg, and S. Wiesner, Phys. Rev. A. \textbf{54}, R1745 (1996).

\bibitem{Sarovar2005} M. Sarovar and G. J. Milburn, Phys. Rev. A \textbf{72}, 012306 (2005).

\bibitem{Paz-Silva2012} G. A. Paz-Silva, A. T. Rezakhani, J. M. Dominy, and D. A. Lidar, Phys. Rev. Lett. \textbf{108}, 080501 (2012).

\bibitem{SupMat} Supplementary Material

\bibitem{Dennis2002} E. Dennis, A. Kitaev, A. Landahl, and J. Preskill, J. Math. Phys. \textbf{43}, 4452 (2002).

\bibitem{Fowler2009} A. G. Fowler, A. M. Stephens, and P. Groszkowski, Phys. Rev. A \textbf{80}, 052312 (2009).

\bibitem{Wang2011} D. S. Wang, A. G. Fowler, and L. C. L. Hollenberg, Phys. Rev. A \textbf{83}, 020302(R) (2011).

\bibitem{Peres1990} A. Peres and A. Ron, Phys. Rev. A \textbf{42}, 5720 (1990).

\end{thebibliography}
\end{document}